\documentclass[prb,twocolumn,preprintnumbers,showpacs]{revtex4}

\usepackage{dcolumn,graphicx,amsmath,amssymb}

\def\degC{$^0$C}

\def\fig#1{Fig.~\ref{#1}}
\def\eq#1{Eq.~(\ref{#1})}

\begin{document}

\title{Effect of annealing on the hyperfine interaction in InAs/GaAs quantum dots}

\author{Michael~Yu.~Petrov,$^{1,}$\cite{e-mail} Ivan~V.~Ignatiev,$^1$ Sergei~V.~Poltavtsev,$^1$ Alex~Greilich,$^2$ Ansgar~Bauschulte,$^2$ Dmitri~R.~Yakovlev,$^{2,3}$ and Manfred~Bayer$^2$}%
\affiliation{$^1$Institute of Physics, St.~Petersburg State University, 198504 St.~Petersburg, Russia\\
$^2$Experimentelle Physik II, Universit\"at Dortmund, D-44227 Dortmund, Germany\\
$^3$A.~F. Ioffe Physico-Technical Institute, RAS, 194021 St. Petersburg, Russia
}%
\date{\today}

\begin{abstract}
The hyperfine interaction of an electron with nuclei in the annealed self-assembled InAs/GaAs quantum dots is theoretically analyzed.
For this purpose, the annealing process, and energy structure of the quantum dots are numerically modeled.
The modeling is verified by comparing the calculated optical transitions with the experimental data on photoluminescence for a set of the annealed quantum dots.
The localization volume of the electron in the ground state and the partial contributions of In, Ga, and As nuclei to the hyperfine interaction are calculated as functions of the annealing temperature.
It is established that the contribution of indium nuclei into the hyperfine interaction predominates up to high annealing temperatures ($T_a=980$\;\degC) when the In content in the quantum dots does not exceed 25\%.
The effect of the nuclear spin fluctuations on the electron spin polarization is numerically modeled.
Effective field of the fluctuations is found to be in good agreement with the available experimental data.
\end{abstract}
\pacs{72.25.Rb, 78.67.Hc, 73.21.La}
\maketitle

\section{Introduction}
Hyperfine interaction of an electron with a nuclear spin ensemble is known to give rise to the most effective mechanism of electron spin relaxation in quantum dots.\cite{GammonPRL01, BraunPRL05, CherbuninPRB07}
Due to limited number of nuclei in a quantum dot (QD) which interact with the electron spin, combination of the randomly oriented nuclear spins leads to non-zero total spin of the nuclear system.\cite{MerkulovPRB02, KhaetskiiPRL02}
This total spin acts on the electron spin as an effective magnetic field, $B_N$, with random magnitude and orientation.
The electron spin rapidly precesses about this fluctuating field that results in a decay of the electron spin polarization in the QD ensemble.
Typical times of the spin decay are of about a fraction of nanosecond for InAs/GaAs QDs.\cite{BraunPRL05}
At the same time, the electron spin relaxation due to other processes such as electron-phonon interaction is a few orders of magnitude longer.\cite{KhaetskiiPRB00, KroutvarNature04}

The hyperfine interaction strength depends on the number of nuclei covered by the electron wave function, thus on the electron localization volume.\cite{MerkulovPRB02}
Scaling the QD effective size we can efficiently control the hyperfine interaction.
A good way to change the QD size is the postgrowth thermal annealing the heterostructure with the QDs at relatively high temperatures.
The annealing causes diffusion of indium from the QDs into the barrier layers that results in decreasing the potential well depths and its enlarging in real space.\cite{LeonPRB98, FafardAPL99}
Besides, the annealing allows one to reduce the number of point defects and, thus, to improve the structure quality\cite{FafardAPL99} that suppresses the defect-related electron spin relaxation mechanism.

In this paper, we theoretically consider the effect of postgrowth annealing on the electron-nuclei hyperfine interaction in the self-assembled InAs/GaAs QDs.
We offer a numerical model of the QD which is based on the real data obtained in experiments with such heterostructures.\cite{CherbuninPRB07, LangbeinPRB04, GreilichPRL06}
This model allows us to simulate the process of annealing and to calculate the distribution profiles of In and Ga atoms over the heterostructure.
Using this model we calculated the energy states of the carriers and the optical transitions in the annealed QDs.
We fitted the parameters of our model using the experimental data on photoluminescence (PL) for a set of the heterostructures with the InAs/GaAs QDs annealed at different temperatures.
This fitting allowed us to model the hyperfine interaction of an electron with the nuclei as a function of annealing temperature of the heterostructure.
Finally, the result of these calculations made it possible to quantitatively describe the electron spin depolarization via the hyperfine interaction with nuclei and suppression this effect by a longitudinal magnetic field.

\section{Quantum Dot Model}
\begin{figure}[t]
\includegraphics*[width=8.6cm]{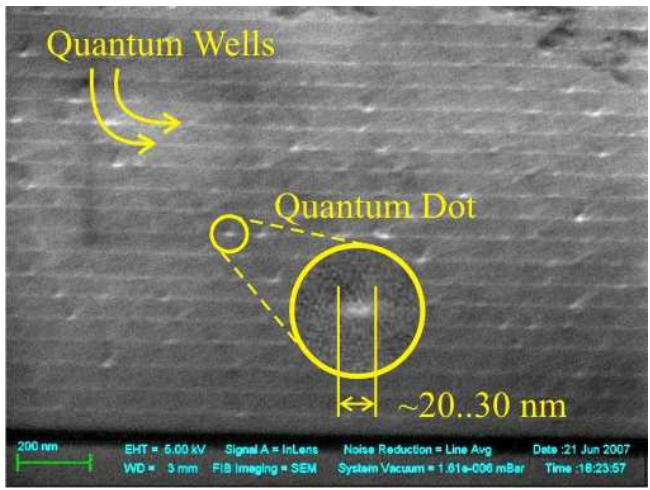}
\caption{SEM image of the cross-section of the heterostructure with unannealed InAs/GaAs QDs.
}\label{figSEMQD}
\end{figure}

Typically the self-assembled InAs/GaAs quantum dots have the truncated pyramid or lens shape with base diameter $d=$~15\,--\,30~nm and height $h=$~5\,--\,15~nm.\cite{BGL}
The size of the QDs depends on the growth parameters, in particular on the nominal thickness of the deposited layer of indium.
Besides, there is a spread of sizes of QDs in the QD ensemble.
To be specific, we consider heterostructure investigated in Refs.~[\onlinecite{CherbuninPRB07, LangbeinPRB04, GreilichPRL06}].
Its cross-section image obtained by scanning electron microscopy (SEM) is shown in \fig{figSEMQD}.
Though the spatial resolution is not high we can estimate the base diameter of the QDs to be of about 20\,--\,25~nm.
A higher spatial resolution can be obtained by transmission electron microscopy, but such data are not available for this heterostructure.

Therefore we use indirect data on the QD shape and size which can be extracted from the analysis of the PL spectra.
As it is well known, the quantization energy of carriers and, therefore, the energies of optical transitions depends on size and shape of a QD.
We choose these parameters to obtain correspondence between the experimentally observed and calculated energies of optical transitions between the lowest and excited states.
The lowest optical transition depends mainly on the size of the QD.
However, the energy distance between the lowest and excited states and relevant optical transition energies depend on the ratio of the height to base diameter of the QD.
We use these circumstances to determine separately the height and the base diameter of the QD under study.
We ignore the statistical spread of sizes of the QDs in ensemble.
Moreover, for simplification of calculations we consider the QD with cylindrical symmetry and smooth bell-like shape.
The QD model is schematically shown in \fig{figModelAndBoundaries}.
The height of the QD $h_{QD}=8$~nm and the base diameter (at height $0.1 h_{QD}$) $d_{QD}=30$~nm.
We put the QD on a thin layer of InAs ($h_{WL}=0.283$~nm) to model the wetting layer (WL) which is inevitably appear when growing the QDs in the Stransky-Krastanow growth mode.

\section{\label{sec:level1}Gallium and Indium Interdiffusion Process due to Annealing of the Heterostructure}
Postgrowth annealing of a heterostructure with self-assembled InAs/GaAs quantum dots leads to indium and gallium interdiffusion.\cite{LeonPRB98, FafardAPL99, MalikAPL97,GunawanPRB05}
Like to other authors,\cite{GunawanPRB05} we consider this process in a model of continuum because the effective scale which we are interested in is much larger then the lattice constant.
Besides we assume the diffusion coefficient to be independent of space coordinates.
So we describe the diffusion by the Fick's law:
\begin{equation}
\frac {\partial x(\boldsymbol{r}, t)} {\partial t} - D \Delta x(\boldsymbol{r}, t) = 0
\text{,}
\label{eqnSecAnnealingFick}
\end{equation}
where $x({\boldsymbol r}, t)$ is a position-dependent function of indium fraction in the In$_x$Ga$_{1-x}$As solid solution forming the QD annealed during the time $t$, and $D$ is the diffusion constant.
The diffusion equation~(\ref{eqnSecAnnealingFick}) should be supplemented with the initial condition on the function $x$.
In our model we assume that the unannealed QD is pure InAs, so that the initial conditions are:
\begin{equation}
x(\boldsymbol{r}, 0) =
\begin {cases}
1,\quad&\text{inside QD/WL} \\
0,\quad&\text{inside the barrier layers.}
\end {cases}
\label{eqnSecAnnealingInitConds}
\end{equation}

\begin{figure}[b]
\includegraphics*[width=8.6cm]{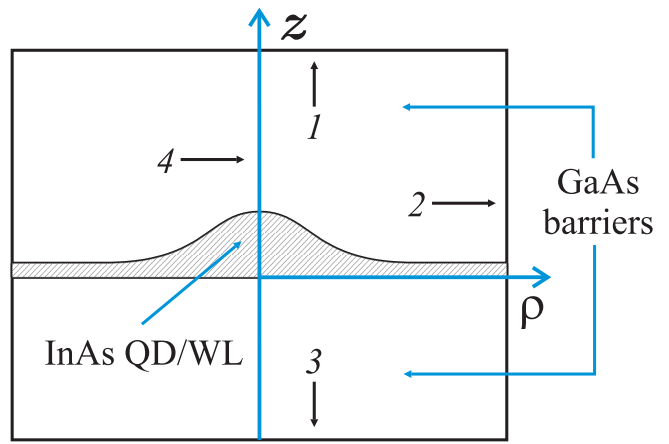}
\caption{Model of the quantum dot: cross-section of the heterostructure and the computational region (see text for details).
}\label{figModelAndBoundaries}
\end{figure}
In the accepted model of cylindrical symmetry of the QDs, we use the cylindrical co-ordinates.
Defining axial axis $z$ to be perpendicular to the plane of the quantum-well layer (WL), and radial axis $\rho$ that lies in the quantum-well plain (see \fig{figModelAndBoundaries}).
In these co-ordinates, the function $x(\boldsymbol{r}, t)$ has got a separable form in azimuthal angle $\varphi$: $x(\boldsymbol{r}, t) = \chi(z,\rho,t)\Phi(\varphi)$.
Because the cylindrical symmetry is conserved during the annealing process, the indium distribution $x$ does not depend on azimuthal angle, i.~e. $\Phi(\varphi) = 1$.
Dividing \eq{eqnSecAnnealingFick} on $D$ and separating the variables, we can write the equation for the function $\chi(z,\rho,t)$:
\begin{equation}
\frac1D \frac\partial{\partial t} \chi - \frac1\rho \frac\partial{\partial \rho} \rho \frac\partial{\partial \rho} \chi - \frac{\partial^2}{\partial z^2} \chi = 0 \text{.}
\label{eqnSecAnnealingDiffusChi}
\end{equation}
The initial conditions for the function $\chi$ coincide with those for $x$ [see \eq{eqnSecAnnealingInitConds}].

We choose the computation region to be a cylinder with the height, $H_{Cyl}=80$~nm, and the diameter, $D_{Cyl}=80$~nm.
As we found, these values are much larger then the diffusion length of indium atoms in InAs/GaAs semiconductors, $<10$~nm, at annealing temperatures up to 1000~\degC.
For the rigorous description of the problem, \eq{eqnSecAnnealingDiffusChi} should be supplemented with appropriate boundary conditions.
The Dirichlet boundary conditions, $\chi=0$, are imposed on boundaries 1 and 3 because these boundaries are far from the QD (see \fig{figModelAndBoundaries}).
We also assume that the indium flux through boundary 2 is zero because, in the neighborhood of this boundary, the In atoms diffuse from the WL in the perpendicular direction which is evident from the local symmetry of the problem.
This assumption can be described by the Neumann boundary condition, $\boldsymbol{n}\cdot \vec{\nabla} \chi = 0$, where $\boldsymbol{n}$ is the outward normal vector to the boundary.
On boundary 4, the Neumann boundary conditions also must be imposed for nulling the diffusion flux across the symmetry axis of the dot.

Using finite element technique, we solve the described diffusion problem with different diffusion constants and the fixed annealing time interval, $t=30$~sec.
We assumed the Arrhenius equation for the temperature dependence of the diffusion constant:
\begin{equation}
D(T_a) = D_0 \exp\left[-\frac{E_A}{kT_a}\right]
\text{,}
\label{eqnSecAnnealingArrhenius}
\end{equation}
where $T_a$ is the annealing temperature, $E_A$ is the activation energy of the interdiffusion process, $k$ is the Boltzmann constant, and $D_0$ is the pre-exponential factor.
We used $E_A$ and $D_0$ as fitting parameters whose values were optimized to get the best agreement between the experimentally measured and calculated PL spectra of the annealed QDs (see Sec.~\ref{sec:level2}).
We found that $E_A = 1.23$~eV and $D_0 = 8.5 \times 10^{-14}$~m$^2$/s.
The obtained value of $E_A$ is approximately three times smaller than reported one in Ref.~[\onlinecite{MalikAPL97}] for annealed InGaAs QDs.
The diffusion length $L_D = \surd[Dt]$, which is usually measured in experiments, is also larger in our calculations ($L_D = 3.6$~nm vs $L_D \simeq 1.5$~nm in Ref.~[\onlinecite{FafardAPL99}] at 900~\degC).
The origin of discrepancy of these quantities is unclear.
We found that variations of parameters of our QD model (size of QD, band-offset, strain energy, see Sec.~\ref{sec:level2}) in the ranges when the calculated optical transitions agree to experiment (see Sec.~\ref{sec:level2}) do not change the activation energy noticeably.

\fig{CandPsi} shows examples of the In distribution over the heterostructure calculated for the QDs unannealed~(a) and annealed at different temperatures~(b\,--\,d).
As seen, annealing of the heterostructure leads to the rapid dissolution of the InAs QD into the GaAs barrier layers.
The average indium content does not exceed 25\% for QDs annealed at temperatures above 980~\degC.
Besides, the QD volume increases with annealing temperature.

\section{Energy Structure of the Annealed Quantum Dots}
\subsection{Potential profiles and effective masses}
Diffusion of indium from the QDs into the barrier layers leads to modification of the profiles of the valence and conduction bands.
Using the calculated indium content $x(\boldsymbol{r})$ for the annealed QD, we modeled the three-dimensional potential profiles as well as the profiles for the effective masses of carriers.

For this purpose, we used a linear approximation for the bandgap of In$_x$Ga$_{1-x}$As using relevant quantities for InAs and GaAs:\cite{Landoldt}
\begin{equation}
E_g(\boldsymbol{r}) = E_g^\text{InAs} \cdot x(\boldsymbol{r}) + E_g^\text{GaAs} \cdot [1 - x(\boldsymbol{r})]
\text{,}\label{EgXDependence}
\end{equation}
where $E_g^\text{InAs} = 0.415$~eV and $E_g^\text{GaAs} = 1.519$~eV are the bandgaps for InAs and GaAs, respectivelly.
Besides, we used the band-offset ratio, $Q_e/Q_h = 76/34$, along with the the data for InAs/GaAs taken from Refs.~[\onlinecite{vdWallePRB89, StierPRB99}].
There is, however, an important problem which complicates this point.
Namely, the large mismatch between the InAs and GaAs lattice constants gives rise to the large built-in strain which considerably affect the potential profiles.\cite{BGL,StierPRB99,GrundmannPRB95,PryorPRB98,CusackPRB96,FonsecaPRB98,LazarenkovaAPL04}

We include this strain in our model in a simple way.
At the begining, let us consider the unannealed QD.
The average value of hydrostatic strain in the InAs/GaAs QDs, 7\,--\,10~\%, which obtained in calculations by different authors using continuum elasticity theory,\cite{StierPRB99,GrundmannPRB95,PryorPRB98,FonsecaPRB98} and valence-force-field Keating model,\cite{StierPRB99,PryorPRB98,CusackPRB96,LazarenkovaAPL04} leads to decreasing the well depth in the conduction band by 350\,--\,500~meV.
Moreover, the strain weakly changes over the QD volume and, therefore, the constant potential approximation gives rise to the good results.\cite{CalifanoPRB00}
We assume in our model the strain-induced energy shift of the conduction band by $\mathcal{E}_{St}^e = 450$~meV, so that the electron confining potential is 390~meV.
Using this value, we are able to correctly describe the optical transition energies observed experimentally (see Sec.~\ref{sec:level2}).

The confinement potential shape in the valence band is more complex than that in the conduction band.\cite{StierPRB99,GrundmannPRB95,PryorPRB98,CusackPRB96,FonsecaPRB98}
However, Califano \emph{et~al.}\cite{CalifanoPRB00} used a simple model of the constant potential for the valence band and determined the potential well depth from comparison of their calculations with the experimentally found energies of optical transitions.
Using similar method, we obtained the strain energy in the valence band $\mathcal{E}_{St}^h = 90$~meV which corresponds to the hole confining potential 175~meV.

The annealing leads to redistribution the indium concentration and, as we believe, to relaxation the built-in strain.
We assume the strain energy linearly depends on the indium concentration in the annealed QDs.
Thus, using the calculated distribution $x(\boldsymbol{r})$, the bandgap (see \eq{EgXDependence}), the band-offset ratio, and the linear strain dependence on $x(\boldsymbol{r})$, we can determine the potential profiles by means of equation:
\begin{equation}
V_{e,h}(\boldsymbol{r}) = Q_{e,h} \left(E_g(\text{GaAs}) - E_g(\boldsymbol{r})\right)
- \mathcal{E}_{St}^{e,h} \cdot x(\boldsymbol{r})
\text{.}\label{eqnSecPotentialsVandCpotentials}
\end{equation}
The cross-sections of potential profiles along $z$-axis thus determined are shown in \fig{figPotentials}.

\begin{figure*}[t]
\includegraphics{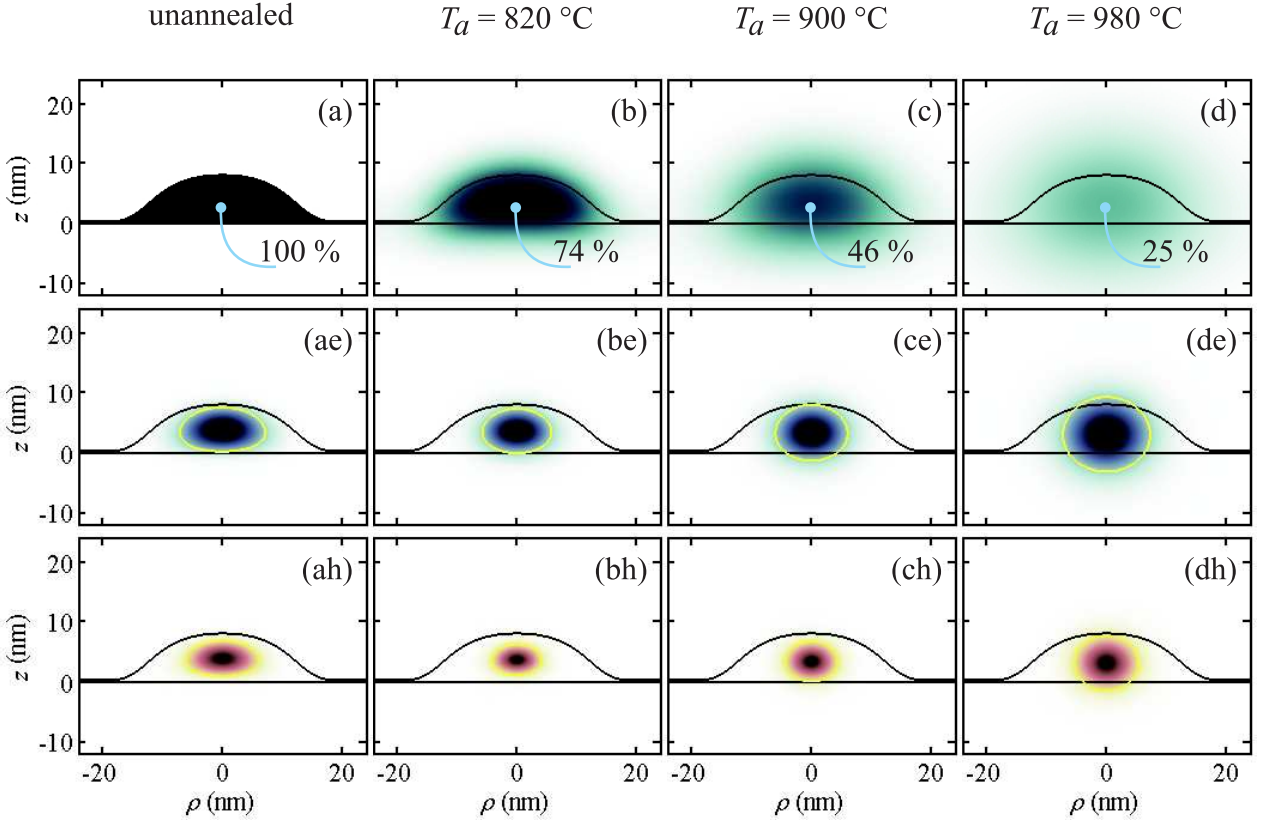}
\caption{(Color online) (a\,--\,d) Cross-section of calculated distributions of indium content $x$ over the heterostructure unannealed and annealed at different temperatures. Highest values of indium concentration are marked in percents for each QDs. Thin black line shows the unannealed QD shape. (ae\,--\,de) Distributions of electron density for the ground state in the unannealed and annealed QDs. The cross-sections of probability density isosurface at level $0.33 \lvert \psi \lvert^2_{\max}$ are shown by thin yellow ellipse. (ah\,--\,dh) Distributions of the ground state hole densities. The notations are similar to those used for the electron density distributions.
}\label{CandPsi}
\end{figure*}
The strain also affects the effective masses of carriers.
In the dot material the compressive stress alters the curvature of the bulk bands, causing the effective masses to differ from the unstrained ones.
We have used the values of electron effective masses in the GaAs barriers and in the strained InAs QD, $m^*_e(\text{GaAs}) = 0.0665 m_0$ (Ref.~[\onlinecite{Landoldt}]), and $m^*_e(\text{InAs}) = 0.04 m_0$ (Ref.~[\onlinecite{CalifanoPRB00}]), respectively.
The hole effective masses used in our calculations are: $m^*_h(\text{GaAs}) = 0.3774 m_0$, and $m^*_h(\text{InAs}) = 0.341 m_0$.\cite{CalifanoPRB00}
Here $m_0$ is the electron mass.
In the annealed QDs, we used the linear dependence of the carrier effective masses on the indium concentration:
\begin{equation}
m^*_{e,h}(\boldsymbol{r}) = m^*_{e,h}(\text{InAs}) \cdot x(\boldsymbol{r}) + m^*_{e,h}(\text{GaAs}) \cdot [1 - x(\boldsymbol{r})]
\text{.}\label{eqnEffectiveMasses}
\end{equation}

\subsection{The electron and hole energy states}
To compute the electron and hole states for the annealed heterostructure, we solve the one-band Schr\"odinger equation in the effective mass approximation:
\begin{equation}
- \frac{\hslash^2}2 \nabla \left[ \frac 1 {m^*(\boldsymbol{r})} \nabla \psi(\boldsymbol{r}) \right]
+ V(\boldsymbol{r}) \psi(\boldsymbol{r})
= E\psi(\boldsymbol{r})
\text{,}
\end{equation}
where $\hslash$ is the Planck's constant divided by $2\pi$, $m^*(\boldsymbol{r})$ and $V(\boldsymbol{r})$ are position-dependent the potential energy and the electron (or hole) effective mass [see Eqs.~(\ref{eqnSecPotentialsVandCpotentials},\ref{eqnEffectiveMasses})], $E$ is the carrier energy, and $\psi(\boldsymbol{r})$ is the envelope wave-function.
Because the annealed QD/WL system in our model has a perfect cylindrical symmetry, we again can write the carrier position in the cylindrical co-ordinate system, $\boldsymbol{r} = (z,\rho,\varphi)$.
We can partially separate the co-ordinates in the total wave-function, $\psi(\boldsymbol{r}) = \eta(z,\rho) \Theta(\varphi)$, and rewrite the Schr\"odinger equation in the cylindrical co-ordinates as:
\begin{equation}
\begin{split}
- \frac{\hslash^2}2 \left[ \frac\partial{\partial z} \left( \frac1{m^*} \frac{\partial\eta}{\partial z} \right)
+ \frac1\rho \frac\partial{\partial\rho} \left( \frac\rho{m^*} \frac{\partial\eta}{\partial\rho}\right) \right] \Theta &-\\
- \frac{\hslash^2}{2m^*} \frac\eta{\rho^2} \frac{\partial^2\Theta}{\partial\varphi^2}
+ V\eta\Theta &= E\eta\Theta
\text{.}
\end{split}
\end{equation}
Dividing this equation by $\eta(z,\rho)\Theta(\varphi)/(m^*\rho^2)$ and rearranging its terms, we come to the two independent equations:
\begin{equation}
\begin{split}
- \frac{\hslash^2}2 \left[ \frac\partial{\partial z} \left( \frac1{m^*} \frac{\partial\eta_n}{\partial z} \right)
+ \frac1\rho \frac\partial{\partial\rho} \left( \frac\rho{m^*} \frac{\partial\eta_n}{\partial\rho} \right) \right] &+\\
+ V\eta_n + \frac{\hslash^2}{2m^*} \frac{n^2}{\rho^2}\eta_n &= E\eta_n
\text{,}
\end{split}
\label{eqnSecEStatesEigenproblem}
\end{equation}
\begin{equation}
\frac{\hslash^2}2 \frac1\Theta \frac{\partial^2\Theta}{\partial\varphi^2} =
- \frac{\hslash^2}2 n^2
\text{.}
\end{equation}
The last equation for $\Theta(\varphi)$ can be solved analytically:
\begin{equation}
\Theta(\varphi) = \frac1{\sqrt{2\pi}} \exp \left[ i n \varphi \right]
\text{,}
\label{eqnSecSchrodAngle}
\end{equation}
where $n$ should be an integer to get one-valued function [$\Theta(2\pi)=\Theta(0)$].

Then, we should discuss the physically substantiated boundary conditions for function $\eta_n(z,\rho)$.
We are only interested in electronic (hole) states confined in the QD.
Therefore, we can impose the Dirichlet boundary conditions on boundaries 1\,--\,3, $\eta_n(z,\rho) = 0$, to provide the wave-function dumping in the barrier layers (see \fig{figModelAndBoundaries}).
Then we consider the boundary conditions on boundary 4 (symmetry axis).
There are two different types of the conditions.\cite{MelnikNTN04}
When $n\geqslant1$, the Dirichlet boundary conditions must be imposed to ensure that the last term in \eq{eqnSecEStatesEigenproblem} does not diverge on symmetry axis, $\rho=0$.
When $n=0$, we employ Neumann boundary conditions, $\boldsymbol{n}\cdot \vec{\nabla} \eta_n(z,\rho) = 0$, to ensure $\vec\nabla\psi$ existing.
Here $\boldsymbol{n}$ is the outward normal vector to the boundary.
Apart from these assumptions, we must impose constraints on function $\eta_n(z,\rho)$,
\begin{equation}
\iint \lvert \eta_n(z,\rho) \lvert^2 \rho \, dz \:\! d\rho = 1
\text{,}
\end{equation}
to satisfy the normalization conditions for the wave-function.

Using finite element technique, we solve eigenvalue problem described by \eq{eqnSecEStatesEigenproblem} for electrons and holes in the QDs, unannealed and annealed at different temperatures.
The annealing leads to a modification the carrier density shape and to an increase the localization area at high annealing temperatures.
Namely, at low annealing temperatures (less than 820~\degC), the shape of the QD is changed a little that almost does not affect the carrier clouds.
However, the changes of potential well shapes promotes some additional localization of the carriers.
The annealing at greater temperatures leads to spherical-like shapes of the carrier density distributions, which corresponds to the spherical-like shape of the indium concentration distribution.
These effects are illustrated in \fig{CandPsi}, where the electron densities (ae\,--de) and hole densities (ah\,--\,dh) for the lowest states in the unannealed and annealed QDs are shown.

\begin{figure}[t]
\includegraphics*[width=8.6cm]{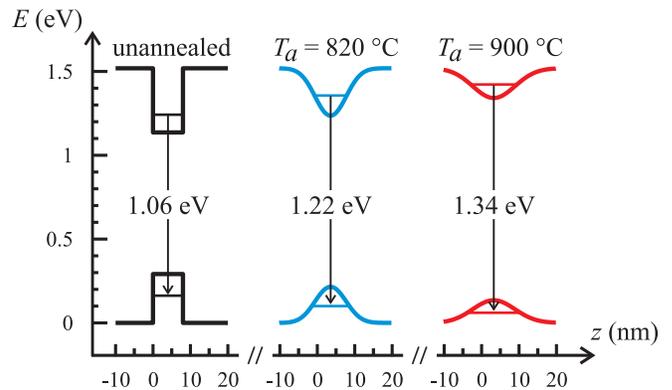}
\caption{(Color online) Potential profiles for the valence and conduction bands along the symmetry axis of problem ($\rho = 0$) for the QDs unannealed and annealed at different temperatures. The energy position of the lowest electron and hole states are shown by thin horizontal lines. The calculated energies of the ground optical transitions are shown in eV.
}\label{figPotentials}
\end{figure}

Also the annealing of the QD leads to an increase in the electron and hole energy levels.
The latter effect is illustrated in \fig{figPotentials}.
As seen, the potential well depth is decreased that provides the decreasing the energy distance between the energy levels and bottom of the potential well.
However, the energy gap between the electron and hole states increases.
This affects the energies of optical transitions in the annealed QDs.

\subsection{\label{sec:level2}Optical transitions and comparison with PL spectra}
The next step of modeling is describing the optical transitions in annealed QDs.
To calculate their energies, we should take into account the energy of the electron-hole Coulomb interaction.
Rigorous solution of this problem is extremely difficult because we must consider the quantum mechanical problem for electron-hole pair in the configuration space with six degrees of freedom.
However, since the localizing potential for carriers in a QD is much larger then the potential due to the Coulomb interaction, we may ignore the Coulomb correlations in the motion of the electron and hole in the QD, and calculate the Coulomb interaction energy, $E_C$, for two fixed spatially distributed charged clouds:\cite{Landau}
\begin{equation}
E_C^{km} = \frac{e^2}{4\pi\varepsilon_0\varepsilon} \iint
\frac{\lvert \psi_e^k(\boldsymbol{r}_e) \lvert^2 \lvert \psi_h^m(\boldsymbol{r}_h) \lvert^2}
{\lvert \boldsymbol{r}_e - \boldsymbol{r}_h \lvert} \ d^3r_e \:\! d^3r_h
\text{,}
\label{eqnPLS2}
\end{equation}
where $e$ is the elementary electronic charge, $\varepsilon_0$ is the vacuum dielectric constant, $\varepsilon$ is the the average dielectric constant in the annealed QD, $\boldsymbol{r}_e$, and $\boldsymbol{r}_h$ are the electron and hole positions, and $\psi_e^k(\boldsymbol{r}_e)$, and $\psi_h^m(\boldsymbol{r}_h)$ are the envelope wave-functions of electron in the $k$th energy state and hole in the $m$th energy state, respectively, which are calculated above in the framework of the one-band problem.
Calculations show that $E_C^{00}$ is of about 21~meV for unannealed QD and decreases down to 15~meV for the QD annealed at 980~\degC.
This energy is slightly smaller when the carriers are in the excited states.

Then the energies of optical transitions are calculated from the simple equation:
\begin{equation}
E_{km} = E_e^k + E_h^m + E_g^x - E_C^{km}
\text{,}
\label{eqnPLS1}
\end{equation}
where $E_e^k$ and $E_h^m$ are the electron and hole energies with respect to In$_x$Ga$_{1-x}$As conduction band bottom and valence band top, respectively, $E_g^x$ is the In$_x$Ga$_{1-x}$As band-gap energy at the QD center.

\begin{figure}[t]
\includegraphics*[width=8.6cm]{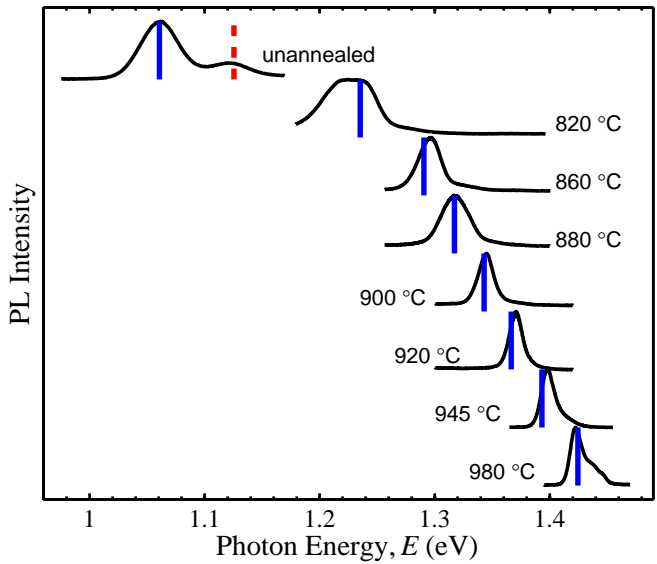}
\caption{Color online) Calculated energies of the lowest optical transitions (solid vertical lines) in comparison with the PL spectra for the QD's unannealed and annealed at different temperatures. The dashed line shows the energy position of the optical transition between the first excited electron and hole states in the unannealed QD.
}\label{figPLSpectra}
\end{figure}
We calculated the optical transitions between the lowest electron and hole energy states as well as between the excited states.
We found that the energy of the lowest optical transition mainly depends on size of the QD.
At the same time, the distance between the lowest and excited transitions is mainly determined by the ratio height/diameter and the shape of the QD.
The strain energies for valence and conduction bands, which we also used as fitting parameters, mainly influence on the shift of optical transition series, and weakly affect the distance between the transition energies.
These facts allow us to uniquely determine the height and the diameter of the QD by comparing the calculated and experimentally measured PL spectra for the unannealed sample.
Then, the annealing was modeled as it is described below in Sec.~\ref{sec:level1}.
We used the interdiffusion activation energy $E_A = 1.23$~eV and the prefactor $D_0 = 8.5\times10^{-14}$~m$^2$/s [see \eq{eqnSecAnnealingArrhenius}] as fitting parameters for all the series of annealed samples.

For comparing with the calculations we measured the PL spectra of a set of the samples using a standard experimental technique.
The spectra are measured at excitation photon energy $E_{ex}=2.54$~eV at temperature $T=1.6$~K.
They are shown in \fig{figPLSpectra}.
It is experimentally and theoretically founded that, the first PL peak corresponds to recombination of the electron and hole in the lowest states, $0e\rightarrow0h$.
The next PL peaks correspond to recombination of an electron and a hole which are in excited states with the same quantum numbers: $1e\rightarrow1h$, $2e\rightarrow2h$, etc.
This interpretation also confirmed by the available experimental data obtained at the high excitation power in magnetic field up to 28~T which are not show here.

As seen from the \fig{figPLSpectra}, the calculated transitions $0e\rightarrow0h$ well reproduce the experiment for all the series.
It should be noted, however, that the very good agreement in energies of the calculated and measured transition $1e\rightarrow1h$ is obtained for the  unannealed sample only.
For the annealed QDs, the calculated energy of the transition is larger by about 20~meV which is comparable with the distance between the $0e\rightarrow0h$ and $1e\rightarrow1h$ transitions.
This disagreement probably related to oversimplifying the strain distribution in our model.
However, because we only interested in the lowest electron states which are involved in the spin relaxation problem under consideration, this disagreement appears to be not very important.

\section{Hyperfine Interaction of the Localized Electron with Nuclei}
\subsection{Effective magnetic field of the nuclear spin fluctuations}
As it is discussed in Introduction, the electron spin polarization is efficiently destroyed in QD ensemble via hyperfine interaction with randomly oriented nuclear spins.
Theoretical justification of the electron spin relaxation mechanism in QDs was reported in Ref.~[\onlinecite{MerkulovPRB02}].
The general idea of this mechanism can be described quantitatively as follows.
The interaction of the electron and nuclear spins is determined by their hyperfine Fermi interaction:
\begin{equation}
\hat{\mathcal H}_{hf} = v_0 \sum_j A^j
\left( \mathbf{\hat S} \cdot \mathbf{\hat I}_j \right)
\left| \psi(\boldsymbol{R}_j) \right|^2
\text{,}
\label{eqnHFI1}
\end{equation}
where $\mathbf{\hat S}$, $\mathbf{\hat I}_j$ are electron and $j$th nucleus spins; $A^j = [16\pi\mu_B\mu_j/(3I_jv_0)] \cdot \lvert u_c(\boldsymbol{R}_j)\lvert^2$ is the hyperfine coupling constant with the $j$th nucleus; $\mu_B$ is the Bohr magneton; $v_0$ is the unit cell volume; $I_j$, $\mu_j$ and $\boldsymbol{R}_j$ are spin magnitude, magnetic moment, and position of the $j$th nucleus; $\psi(\boldsymbol{R}_j)$ and $u_c(\boldsymbol{R}_j)$ are the electron envelope wave-function and the Bloch function at the nuclear site.

Due to the limited number of nuclei in a QD which interact with the electron spin, random orientation of the nuclear spins give non-zero total spin which has a magnitude fluctuating from dot to dot.
The total nuclear spin acts on the electron spin as an effective nuclear hyperfine magnetic field, $\boldsymbol{B}_N$.
We consider the non-polarized and not interacting nuclear spins, since the magnitude and orientation of the effective field are random and can be described by the normal distribution:\cite{MerkulovPRB02}
\begin{equation}
w_B = \frac1{(\sqrt{2\pi}\Delta_B)^3} \exp\left[-\frac{(\boldsymbol{B}_N)^2}{2\Delta_B^2}\right]
\text{,}
\label{eqnHFI2}
\end{equation}
with variance $\Delta_B$ determined by:
\begin{equation}
\Delta_B^2 = \frac13 \sum_j I_j (I_j + 1) b_j^2
\text{,}
\label{eqnHFI3}
\end{equation}
where $b_j = [v_0/(\mu_B{\mathrm g}_e)] A^j \lvert\psi(\boldsymbol{R}_j)\lvert^2$ is the effective magnetic field of a single nuclear spin acting on the electron.
Here ${\mathrm g}_e$ is the electron g-factor.
Unlike paper by Merkulov \emph{et al.},\cite{MerkulovPRB02} we defined variance $\Delta_B$ [see \eq{eqnHFI3}] so that it approximately corresponds to half width at half maximum of normal distribution given by \eq{eqnHFI2}.

In the InGaAs QDs, there are three types of nuclei.
We consider the fluctuating field as a sum of three independent contributions with normal distribution of each of them.
The total variance squared known to be calculated as a sum of variance squares of variate independent contributions, and, therefore, we can write:
\begin{equation}
\Delta_B^2 = (\Delta_B^\text{In})^2 + (\Delta_B^\text{Ga})^2 +(\Delta_B^\text{As})^2
\text{,}
\label{eqnHFI4}
\end{equation}
where $\Delta_B^\text{In}$, $\Delta_B^\text{Ga}$, and $\Delta_B^\text{As}$ are the partial contributions of each types of nuclei:
\begin{equation}
(\Delta_B^\xi)^2 = \frac13 \sum_{j_{\xi}} I_{j_{\xi}} (I_{j_{\xi}} + 1) b_{j_{\xi}}^2
\text{,}
\quad \xi=\text{In,~Ga,~As}
\text{.}
\label{eqnHFI5}
\end{equation}
Here the sum goes only over each types of nuclei in crystal lattice.
Supposing that the electron envelope wave-function is constant over the crystal unit cell and replacing the sum over unit cells by the heterostructure volume integration, we obtain:
\begin{eqnarray}
\begin{split}
(\Delta_B^\text{In})^2 &= C_\text{In} \int \lvert\psi(\boldsymbol{r})\lvert^4 x(\boldsymbol{r}) d^3r
\text{,}\\
\label{eqnHFI6}
(\Delta_B^\text{Ga})^2 &= C_\text{Ga} \int \lvert\psi(\boldsymbol{r})\lvert^4 [1 - x(\boldsymbol{r})] d^3r
\text{,}\\
(\Delta_B^\text{As})^2 &= C_\text{As} \int \lvert\psi(\boldsymbol{r})\lvert^4 d^3r
\text{,}
\end{split}
\end{eqnarray}
where $x(\boldsymbol{r})$ is the indium fraction in In$_x$Ga$_{1-x}$As solid solution forming the QD, and constants $C_\xi$ are determined by:
\begin{equation}
C_\xi = \frac13 I_\xi(I_\xi+1) \frac{A_\xi^2}{(\mu_B{\mathrm g}_e)^2} v_0\text{,} \quad \xi=\text{In,~Ga,~As}
\text{.}
\label{eqnHFI6}
\end{equation}
Actually $x(\boldsymbol{r})$ determines the probability to find the indium nucleus in the position $\boldsymbol{r}$.

Next, we introduce the new parameter, the average effective indium fraction in the QD:
\begin{equation}
\overline{\mathbf{x}} = \frac{\int \lvert\psi(\boldsymbol{r})\lvert^4 x(\boldsymbol{r}) d^3r} {\int \lvert\psi(\boldsymbol{r})\lvert^4 d^3r}
\text{.}
\label{eqnHFI7}
\end{equation}
Using the electron localization volume, defined in Ref.~[\onlinecite{MerkulovPRB02}]:
\begin{equation}
V_L = \left( \int \lvert\psi(\boldsymbol{r})\lvert^4 d^3r \right)^{-1}
\text{,}
\label{eqnHFI8}
\end{equation}
we come to the final expression for the total variance of the effective nuclear field:
\begin{equation}
\Delta_B^2 = \frac1{V_L} \left( C_\text{In}\overline{\mathbf{x}} + C_\text{Ga}(1-\overline{\mathbf{x}}) + C_\text{As} \right)
\text{.}
\label{eqnHFI8}
\end{equation}

We calculated the effective indium fraction and the electron localization volume as a function of the QD annealing temperature.
Then, using \eq{eqnHFI8}, we calculated the total variance and the partial contributions of each types of nuclei.
In these calculations, we used the semiconductor parameters taken from Ref.~[\onlinecite{BraunPRB06}]:\\
\noindent \begin{tabular}{lccc}
\tabularnewline
\hline
\hline
Nuclei & In & Ga$^*$ & As \tabularnewline
\hline
Nuclear spin $I$ & 9/2 & 3/2 & 3/2 \tabularnewline
Hyperfine constant $A\text{~($\mu$eV)}$ & 56 & 42 & 46 \tabularnewline
\hline
\hline
\multicolumn{2}{l} {$^*$\rule{0pt}{11pt}\footnotesize
    Average between $^{69}\text{Ga}$ and $^{71}\text{Ga}$.}\tabularnewline
\tabularnewline
\end{tabular}

\begin{figure}[t]
\includegraphics*[width=8.6cm]{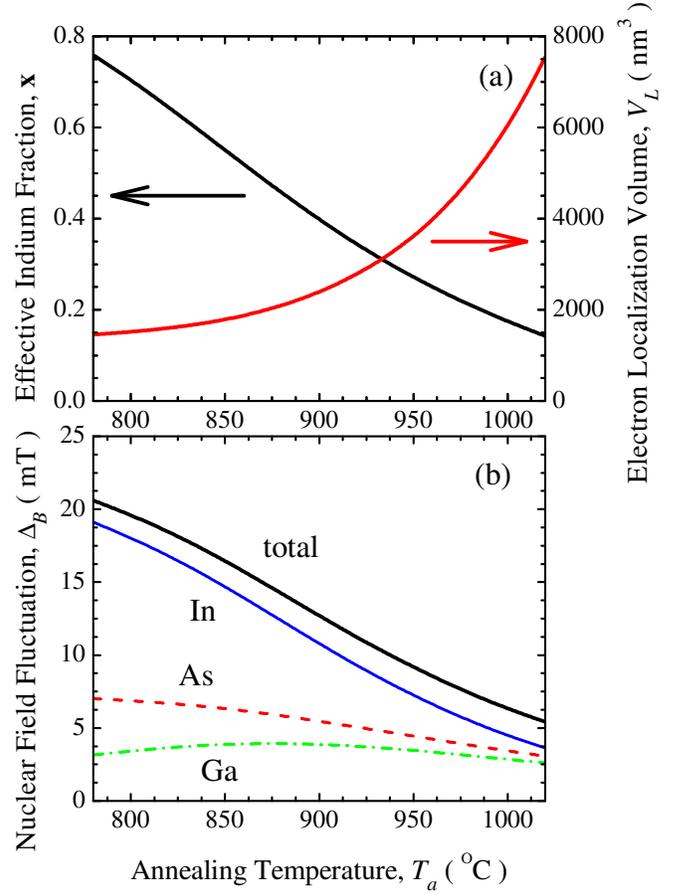}
\caption{(Color online) \textbf{(a)} Dependencies of the electron localization volume (red) and of the average effective indium fraction in the QD (black) on the annealing temperature; \textbf{(b)} The total nuclear field variance and the partial contributions of each type of nuclei as a functions of the QD annealing temperature.
}\label{figDeltaBandVL}
\end{figure}
Results of the calculations are shown in \fig{figDeltaBandVL}.
As seen, the electron localization volume increases with temperature which is due to decrease of the localizing potential depth in QD and the increase of the QD size.
Also, the dissolution of the QD in the barriers leads to decreasing of the effective indium concentration $\overline{\mathbf x}$.
Two these effects result in the decrease of the effective nuclear field variance.
From physical point of view, the variance decrease is due to (i) decrease of contribution of each nucleus to hyperfine interaction because of reducing of the electron density at the nucleus and (ii) averaging of the contributions over increasing number of nuclei.
As seen from \fig{figDeltaBandVL}~(b), the gallium contribution to the dispersion slightly increases with annealing up to temperature $T_a=850$~\degC\ which is caused by the gallium diffusion into the QD.
However, this effect does not influence upon behavior of the total variance because the indium contribution is dominating at all annealing temperatures due to the large indium nuclear spin $I_\text{In}=9/2$.

\subsection{Suppression of the nuclear spin fluctuations by the external magnetic field}
\begin{figure}[t]
\includegraphics*[width=8.6cm]{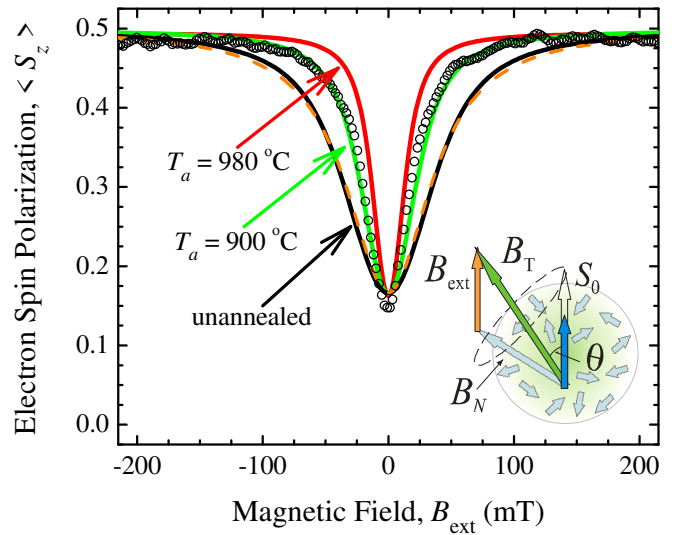}
\caption{(Color online) Calculated external magnetic field dependencies of electron spin polarization in the QDs unannealed and annealed at different temperatures (solid lines) comparing with experiment\cite{CherbuninThanks} (circles). Dashed line is the approximation of calculations by the Lorentz function [\eq{eqnHFI_Lorentz}] for the unannealed sample.
Right inset shows the mechanism of suppression of the nuclear spin fluctuation effect by an external magnetic field.
}\label{figDips}
\end{figure}

The electron spin relaxation caused by the nuclear fluctuating field can be suppressed by applying the external magnetic field.\cite{MerkulovPRB02}
In the presence of the external magnetic field, the electron spin precesses about the total field, $\boldsymbol{B}_\mathrm{T} = \boldsymbol{B}_\mathrm{ext} + \boldsymbol{B}_N$ (see inset in \fig{figDips}).
At sufficiently large external field, the nuclear spin fluctuations almost do not contribute to the total field and, therefore, the electron spin polarization does not decay.

Let us analyze this effect in more detail.
We consider behavior of the projection of the electron spin ($z$-projection) onto the axis of the optical excitation which is typically measured in experiment.
The direction of the external magnetic field (longitudinal field) coincides with $z$-axis.
The electron spin $z$-projection should be averaged over many periods of the spin precession about the total field and over the QD ensemble.
The time averaging allows us to calculate the constant component measured in such experiments.
Although all the QDs in the ensemble are identical in our model, the averaging over the ensemble arises because of the random magnitude and orientation of the nuclear spin fluctuations.
The spin projection is calculated as:
\begin{equation}
< S_z > = \iiint_{-\infty}^\infty S_z(B_\mathrm{ext}) w_B(\boldsymbol{B}_N) d^3B_N
\label{eqnHFI_SzAvg}
\text{.}
\end{equation}
Here, the probability distribution $w_B(\boldsymbol{B}_N)$ is given by \eq{eqnHFI2} and $S_z(B_\mathrm{ext})$ is:
\begin{eqnarray}
\begin{split}
S_z (B_\mathrm{ext}) &= S_0 [\cos \theta(B_\mathrm{ext})]^2 =\\
&= S_0 \frac {({B_N}_z + B_\mathrm{ext})^2} {({B_N}_z + B_\mathrm{ext})^2 + {B_N}_x^2 + {B_N}_y^2}
\text{,}
\end{split}
\label{eqnHFI_Sz}
\end{eqnarray}
where $S_0$ is the initial electron spin polarization (we assume $S_0=1/2$), and $\theta$ is the angle between $z$-axis and the total magnetic field direction (see inset in \fig{figDips}).
Note, that \eq{eqnHFI_SzAvg} is the averaging over the ensemble of the electron spin projection and \eq{eqnHFI_Sz} is a result of the time averaging of the electron spin precession.
Fast precession of the electron spin about $B_\mathrm{T}$ conserves only the projection of the initial spin $S_0$ onto the direction of $B_\mathrm{T}$ so that $S=S_0\cos\theta$.
Therefore, the measurable quantity is $S_z = S\cos\theta = S_0 \cos^2\theta$.

Using the values of $\Delta_B$ found above [see \fig{figDeltaBandVL}~(b)], we calculated the dependencies of $S_z$ as functions of the external magnetic field for the QDs annealed at different temperatures.
The results are shown in \fig{figDips} (solid lines).
As seen, each calculated dependence has a dip around zero external field which is due to the depolarization of the electron spin by the effective nuclear field.
This depolarization is incomplete, namely $(1/3) S_0$ is still conserved.
The result can be understood from a simple consideration.
We may replace the arbitrary oriented nuclear fields by three components oriented along the $x$, $y$, and $z$ co-ordinate axes with equal probabilities, $P = 1/3$, and write the magnetic-field-dependent value of the electron spin upon action of these components:
\begin{eqnarray}
\begin{split}
B_N \parallel x,y: \qquad S_z &= S_0 \frac{B_\mathrm{ext}^2}{B_\mathrm{ext}^2 + B_N^2}
\text{,}\\
B_N \parallel \phantom{x,}\,z:   \qquad S_z &= S_0
\text{.}
\end{split}
\end{eqnarray}
As seen, the nuclear fields along $x$ and $y$ axes totally depolarize the electron spin at zero external magnetic field, and the nuclear field along $z$-axis keeps its initial value.
Assuming that all the components of nuclear field have identical average variances, which can be defined as $<{B_N^2}_{x,y,z}> = (2\Delta_B)^2$, the average electron spin polarization can be written:
\begin{equation}
\bar{S}_z = S_0 \left( \frac13 + \frac23\frac{B_\mathrm{ext}^2}{B_\mathrm{ext}^2 + (2\Delta_B)^2} \right)
\text{.}
\end{equation}
Rearranging terms in this equation gives the magnetic field dependence with the Lorentz-like dip:
\begin{equation}
\bar{S}_z = S_0 \left( 1 - \frac{2/3}{1 + [B_\mathrm{ext}/(2\Delta_B)]^2} \right)
\text{.}\label{eqnHFI_Lorentz}
\end{equation}
As seen from \fig{figDips}, the calculated field dependencies of $<S_z>$ can be perfectly approximated, in the average, by this Lorentz function.

The dip widths at half minima are of several tens of milli-Tesla and decrease with the annealing temperature.
That reflects the decreasing of the hyperfine interaction strength in conformity with the above discussion.
We can characterize the interaction strength by the averaged nuclear field, $\bar{B}_N \simeq 2\Delta_B$, which we define as a half width at half minimum (HWHM) of the dip.
This means that $\bar{B}_N$ and its dependence on the QD annealing temperature can be obtained from those quantities for $\Delta_B$ shown in \fig{figDeltaBandVL}~(b).

The electron spin polarization in the QDs annealed at 900~\degC\ was experimentally studied in the paper by Cherbunin, \emph{et al}.\cite{CherbuninPRB07}
They have measured the circular PL polarization of the singly negatively charged QDs and have found that the polarization is closely related to the spin orientation of the resident electrons.
The magnetic field dependence of the PL polarization is found to reveal a dip around $B_\text{ext}=0$.
One of the experimental curves measured at the excitation density 4.4~W/cm$^2$ is shown by circles in \fig{figDips}.
As seen, the curve is very similar to those calculated theoretically.
A small disagreement is partially related to the experimentally found dependence of the HWHM on the excitation density.\cite{CherbuninPRB07}
With decreasing the excitation density down to $P=0.5\;\text{V/cm}^2$, the HWHM increased up to approximately 30~mT.
The calculated value of HWHM, $\bar{B}_N \simeq 27$~mT, for $T_a=900$~\degC\ is close to experimental one.

\section{Conclusion}
Theoretical modeling of InAs/GaAs QDs allowed us to simulate the effect of nuclear spin fluctuations on the electron spin polarization.
We determined the electron localization volume and the effective indium fraction in the QDs for different annealing temperatures.
Due to interdiffusion of In and Ga during the annealing process, the QD size increases and, correspondingly, the electron localization volume considerably increases (from $\sim1700$~nm$^3$ for unannealed QD up to $\sim4900$~nm$^3$ for the QD annealed at 980~\degC).
At the same time, the dissolution of the QD results in decreasing the effective indium concentration in the QD.
We calculated the partial contributions of the indium, gallium, and arsenic nuclei to the effective magnetic field of the nuclear spin fluctuations and found that the hyperfine interaction is determined mainly by the indium contribution.
The effect of the fluctuations decreases with the annealing temperature due to (i) the increasing number of nuclei interacting with electron and (ii) the decreasing indium concentration.
The average magnitude of the effective hyperfine field decreases from $\sim41$~mT down to $\sim15$~mT with the annealing temperature up to 980~\degC.
Finally, we modeled the suppression of the nuclear spin fluctuation effect in the longitudinal magnetic field.
The calculated dip of the electron spin polarization is very similar to that observed in the experiment.\cite{CherbuninPRB07}

\section*{ACKNOWLEDGMENTS}
The authors thanks R.~V. Cherbunin, I.~Ya. Gerlovin, G.~G. Kozlov, and I.~A. Yugova for fruitful discussions.
This work has been supported in part by the Russian Ministry of Science and Education (grant RNP.2.1.1.362) and by Russian Foundation for Basic Research.


\end{document}